\documentclass[preprint,12pt]{elsart}

\usepackage{epsfig}

\usepackage{amssymb}

\usepackage[nodots,nocompress]{numcompress}

\journal{Information Processing Letters}

\begin{document}

\begin{frontmatter}



\title{Determining All Maximum Uniquely Restricted Matching in Bipartite Graphs}


\author{Guohun Zhu}

\address{Guilin University of Electronic Technology}
\address{No.1 Jinji Road, Guilin, China, 541004 }

\begin{abstract}
The approach mapping from a matching of  bipartite graphs  to  digraphs has been successfully used for forcing set problem, in this paper, it is extended to uniquely restricted matching problem. We show to  determine a uniquely restricted matching in a bipartite graph is equivalent to  recognition a acyclic digraph. Based on these results, it proves that determine the bipartite graphs with all maximum matching are uniquely restricted is polynomial time. This answers an open question of 
Levit and Mandrescu(Discrete Applied Mathematics 132(2004) 163--164). 
\end{abstract}

\begin{keyword}

Bipartite graph \sep  Directed cycle \sep BD-mapping \sep Uniquely restricted matching 
\end{keyword}
\end{frontmatter}


\section{Introduction}

Let $G=(X,Y; E)$ be a bipartite graph, a set of edges $M \subset(X,Y)$ is a matching if 
no two edges of $M$ share a common vertex. 
A matching $M$ is uniquely restricted if its saturated vertices
induce a subgraph which has a unique perfect matching and denotes as   $M_{ur}$. 
A subset edges $S \subset M$ is a forcing set for a matching $M$ if $S$ is in no other perfect matching of $G$.
Let us denote the subgraph induced by the dedges of $M$(laso known as the saturated vertices) as $G[M]$, and  name all of the vetices not saturated by $M$ as free vertex set $V_f$.

Maximum matching problems are well known problem in graph theory
and are proved to  be solved in polynomial time\cite{garow1976}. 
But many restricted maximum matching problems are
NP-complete, for example, Finding the maximum $M_{ur}$ is NP-complete  in bipartite graphs \cite{Golumbic2001},
the smallest forcing set problem is also NP-complete in cubic bipartite graphs \cite{peter2004}

On the positive side, it is proved that the determine a matching is uniquely
restricted  in bipartite graph could be recognized in $O(M+E)$
 \cite{Golumbic2001}. There exists a polynomial time algorithm 
to determine the $M_{ur}$ if $G$ is unicycle graph\cite{Levit2005}. 
And \cite{Levit2004}  also shown that unique
perfect matching bipartite graph could be find in polynomial time.
At the end of \cite{Levit2004}, they raised an open problem as follows:

\begin{description}
\item[Problem.] how to recognize that all maximum matching in a bipartite graph are uniquely restricted?
\end{description}

In this paper we will answer this question in two steps. Firstly, it shows a mapping from bipartite graph 
to digraph, and then it gives a necessary and sufficient condition on a uniquely restricted matching in bipartite graphs 
is  equaivlent to the acyclic digraph. Secondly, 
it proves to determine all the maximum matching uniquely restricted or not 
is eqaivlent to find no more than two path between two vertices.
In addition, it shows that uniquely perfect matching in bipartite graphs is as simple as recoginze the an acyclic digraph.

\section{Illustration the main technology}

The main technonlogy in this paper have a successful implementation on  finding forcing set problem in \cite{peter2004}.  But firstly, let us repeat the theorem in \cite{Golumbic2001}.

\begin{thm}
\label{Golumbic_1} 
\cite{Golumbic2001}
A matching  $M$ of a graph is uniquely restricted
if and only if $M$ is alternating cycle-free.
\end{thm}

Then let us define a mapping from $G[M]$ of a bipartite graph to a digraph $D$ and named as
$BD$-mapping in this paper, this mapping is much more clearly than the definition on page 292 of   \cite{pachter1998} and definition on page 3 of  \cite{peter2004} which denotes by $D(G,M)$.

\begin{defn}
\label{def_1}
Given a matching $M$ of bipartite graph $G(X,Y;E)$, a $BD$-mapping digraph $D(V,A)$ of $G$ defines as  follows
\begin{equation}
 V=\{ x | (x,y) \in M \}.
\end{equation}
\begin{equation}
\label{edge2arc2}
 A=\{ <x_1,x_2> | (x_1,x_{1}^{\prime}) \in M \wedge
(x_2,x_{2}^{\prime}) \in M  \wedge (x1_,x_{2}^{\prime}) \in E-M
 \}.
\end{equation}
\end{defn}

It is easy to observe that follows theorem could be equivalent to
the theorem~\ref{Golumbic_1}.

\begin{thm}
\label{main_thm1}
 Let $D$ be $BD$-mapping digraph of a matching $M$ in bipartite graph with $n >2$ vertices,
$M$ is uniquely matching in $G$ if and only if $D(G,M)$ is acyclic.
\end{thm}

\begin{pf}
Suppose that $D$ is acyclic digraph, every vertex $x$ on $D$ could
be divided into a pair of vertex $<x,x^\prime>$ and become a new
directed graph $D^\prime$, which is also a acyclic digraph and
without alternative cycle. Moreover, there has a matching $M$
include number of $|D|$ edges. Therefore $M$ is a uniquely restricted matching.

On the other hand, assume $M$ is a uniquely restricted matching but the
$BD$-mapping $D(G,M)$ include at least one cycle $C$,
where $C=\{ <x_1,x_2>,<x_2,x_1> \}$  or $C=\{ <x_1,x_2>,<x_2,x_3>
\ldots <x_k, x_1> \}$ when $k \geq 3$. Then according to
equation~\ref{edge2arc2} of definition~\ref{def_1}, there exists $ (x_1, x_1^{\prime}), (x_2,
x_2^{\prime}), (x_1, x_2^{\prime})$ and $(x_2,x_1^{\prime})$ are in
$E$, or  $ (x_1, x_1^{\prime}), (x_2, x_2^{\prime}), \ldots, (x_k,
x_k^{\prime}) \in M$,  and $M^\prime=(x_1, x_2^{\prime}, \ldots,
(x_k,x_1^{\prime}) \in E-M$. Therefor, $M \cup M^\prime$ has a even cycle,
this contradict to the theorem~\ref{Golumbic_1}.
\end{pf}

\begin{rem}
Theorem~\ref{main_thm1} is very similar the proposition $3$ in \cite{peter2004}.
\end{rem}
\begin{prop}
\cite{peter2004}
\label{prop1}
Let $G$ be a bipartite graph, $M$ is a perfect matching of $G$, 
and $S \subset M$ is a  focing set of $M$ if and only if $D(G,M) \setminus S$  is an 
acyclic digraph.
\end{prop}

In fact, Theorem~\ref{main_thm1} can deduce the known results that  

\begin{cor}
The bipartite graph $G$ with uniquley perfect matching $M$ has a forcing matching number $0$ 
\end{cor}
\begin{pf}
Since $D(G,M)$ with uniquely perfect matching is acyclic graph, the $S$ in proposition~\ref{prop1}  is empty set.
\end{pf}

Let us give an example to show a bipartite graph $G$ and the $D(G,M)$ in Fig.$1$ to end of this section.

\begin{figure}[htbp]
\label{fig1}
\centerline{\includegraphics[scale=1]{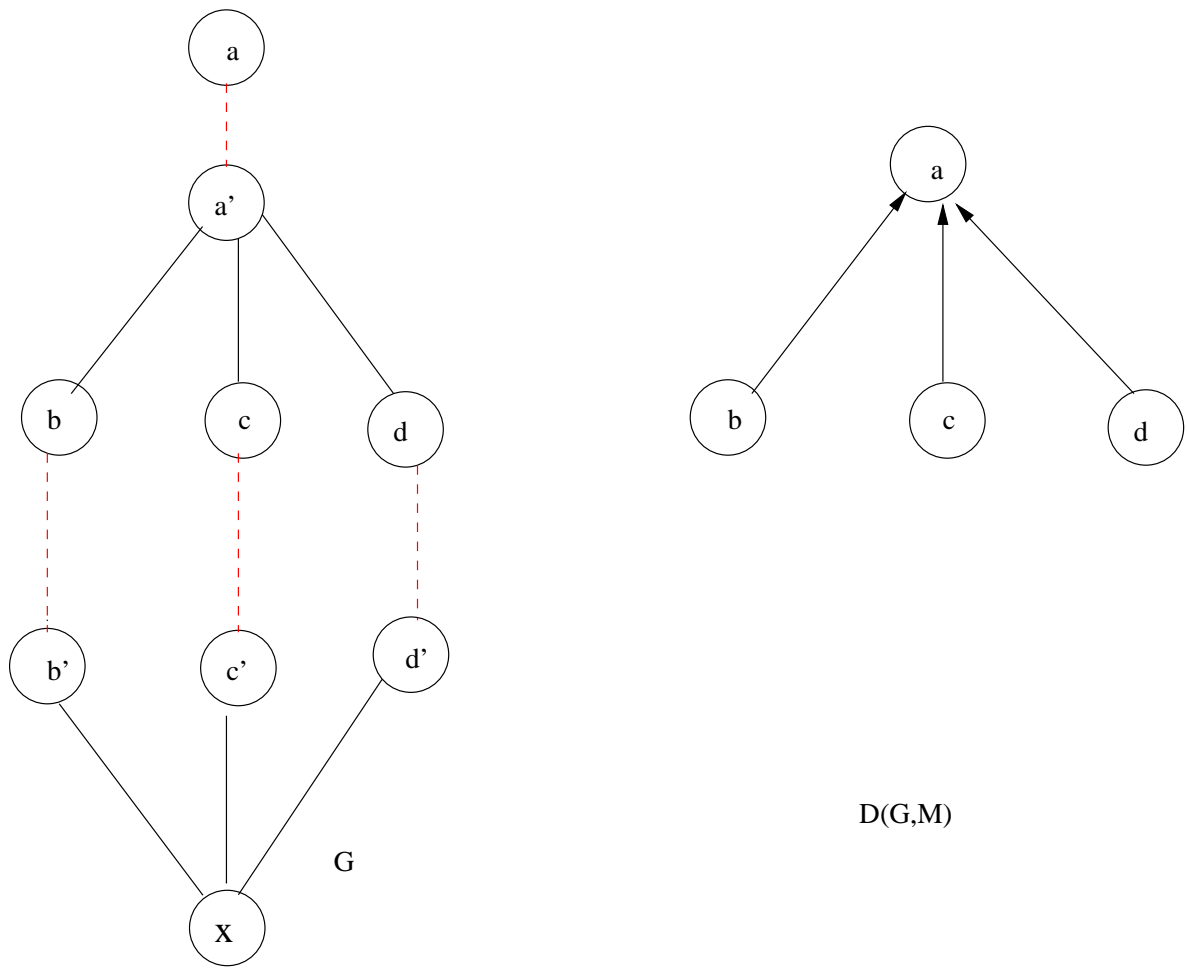}} 
\caption{ The left is bipartite graph $G$ and the right is the $D(G,M)$}
\end{figure}

\section{The complexity of uniquley restricted  perfect matching}


\cite{Levit2004}has proved that if and only all of local
maximum stable set is a greedoid, then a bipartite graph has a
unique perfect matching. However, how to recognize all of local
maximum stable set are greediod is equivalent to the problem of all
maximum maching are uniquely restricted according to the theorem 3.3 in
\cite{Levit2004}. This section would give a more efficient algorithm
to determine the unique perfect matching.

It is easy to obervious to obtain the  following theorem:
\begin{thm}
\label{main_thm2}
A matching  $M$ of a bipartite graph is perfect
uniquely restricted if and only if $D(G,M)$ is acyclic and
$|D(G,M)|=\frac{n}{2}$.
\end{thm}

Based on theorem $7$
an algorithm shows in Algorithm 1.
\begin{algorithm}
\caption{Determine unique pefect matching}
\label{alg1}
\begin{center}
    \fbox{
      \begin{minipage}{\textwidth}
      \begin{tabbing}
{\bf Input} : A bipartite graph $G(X,Y; E)$ and a matching $M \in E$; \\
{\bf Output}: unique perfect maching if it has, otherwise return non unique perfect matching.\\
\noindent{\bf begin}	\\
(1).Generate a BD-mapping graph $D(V; A)=f(G[M])$.	\\
(2)  {\bf if }  $M$ is not perfect  then 			\\
(3)  \ \ \ \ \ \ return non unique perfect matching\\
(4)   {\bf    else}   			\\
(5)   \ \ {\bf if} $D$ is acyclic then return unique perfect matching 	\\
(6)   {\bf    else return non unique perfect matching}   			\\
\ \    {\bf endif}	\\
\noindent{\bf end}	
     \end{tabbing}
     \end{minipage}
    }
  \end{center}
\end{algorithm}

\begin{figure}[htbp]
\label{fig2}
\centerline{\includegraphics[scale=1]{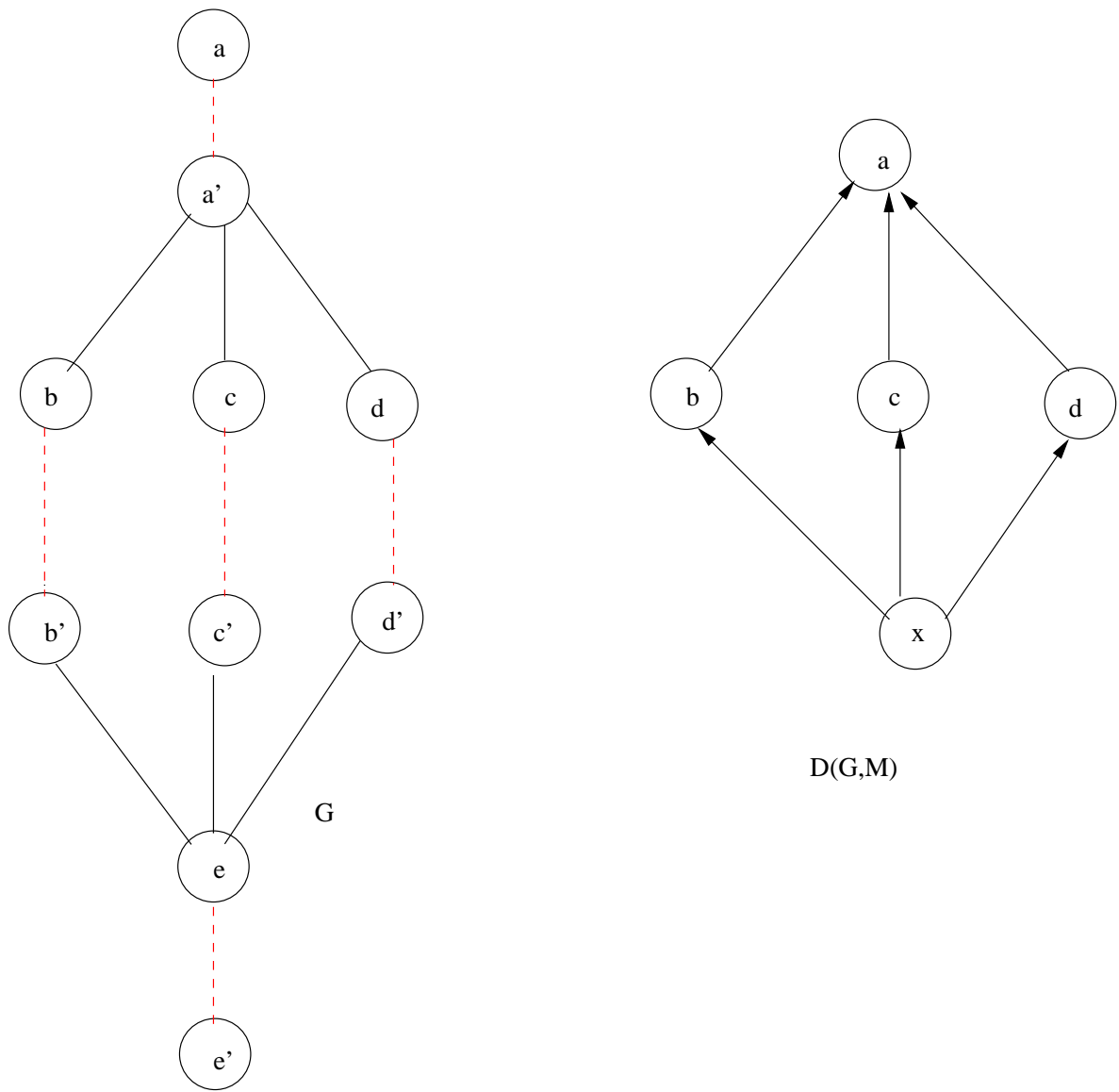}}
\caption{ $G$ with uniquely prefect matching }
\end{figure}

It is clearly to see that the example in Fig.$2$ is a uniquly restricted perfect matching.
  Now let us consider the Fig.$1$ again. 
It is  clearly that  G in Fig.$1$  is contains a unique maximum matching $\{(b,b^\prime),(c,c^\prime),(d,d^\prime),(a^\prime, a)\}$. 
Notice that $G$ also contains a maximum matching $\{(x,b^\prime),(b,a^\prime),(c,c^\prime),(d,d^\prime)\}$ is not uniquely restricted.  
Thus the  theorem~\ref{main_thm1} is only necessary condition for all maximum matching are uniquely restricted.

\section{Determine all uniquely restricted maximum matching in polynomial time}
In first glance, greedy algorithms can apply into determing all uniquely restricted maximum matching 
by remove the node with degree $1$. Unfortunately, the worst case could  be exponent. 

For example, let consider the bipartite graph in Fig.$3$. 
\begin{figure}[htbp]
\centerline{\includegraphics[scale=0.5]{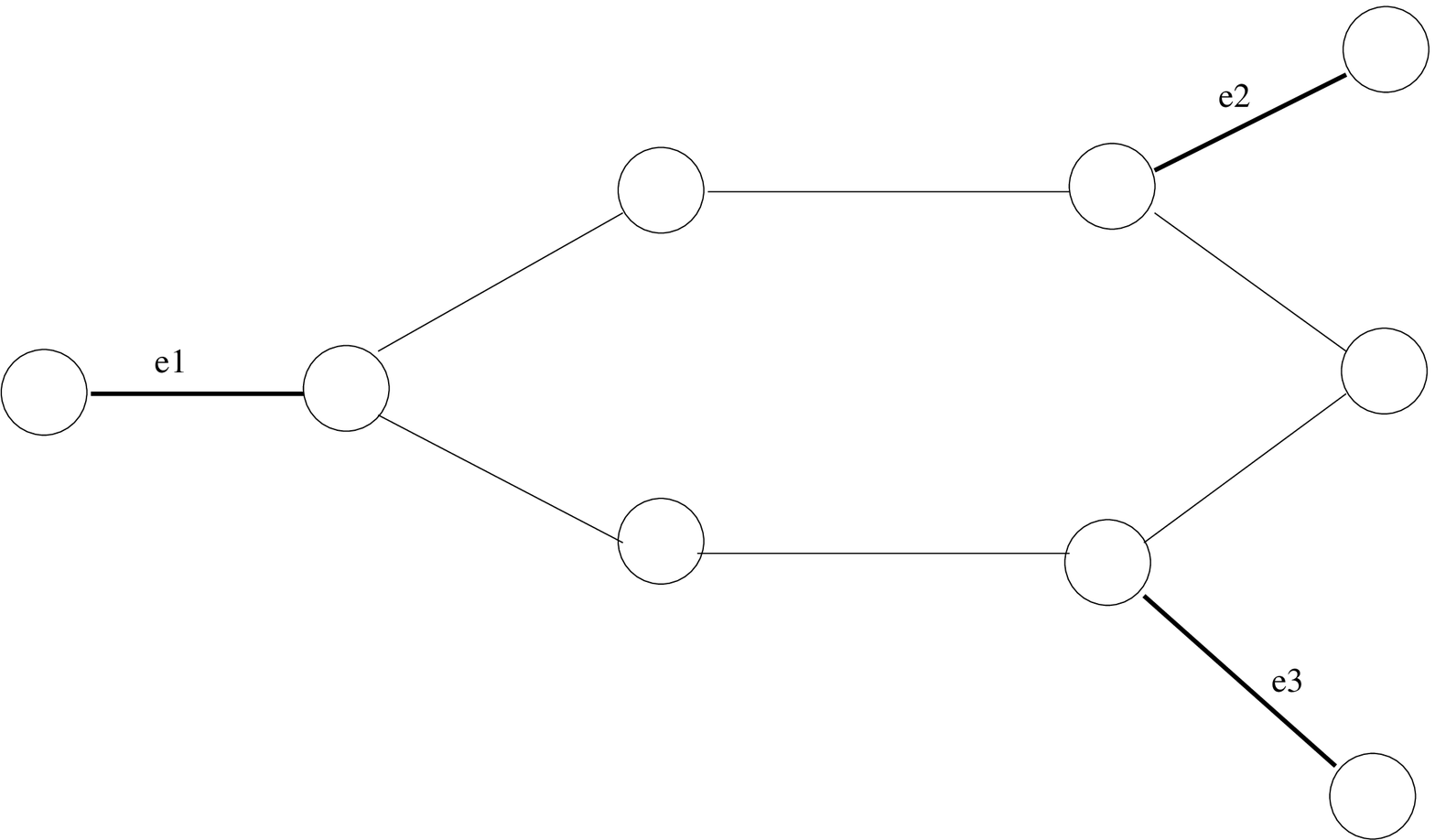}}
\label{fig3}
\caption{Not all maximum matching are uniquely restricted, even remove edge $e1$, there exists $|M_{ur}|=3$ and $|M|=3$.}
\end{figure}
It is need to remove $3$ edges $e1,e2,e3$, then any maximum matching is not uniquely restricted.
But for another example in Fig~\ref{fig4}, 
\begin{figure}[htbp]
\centerline{\includegraphics[scale=0.5]{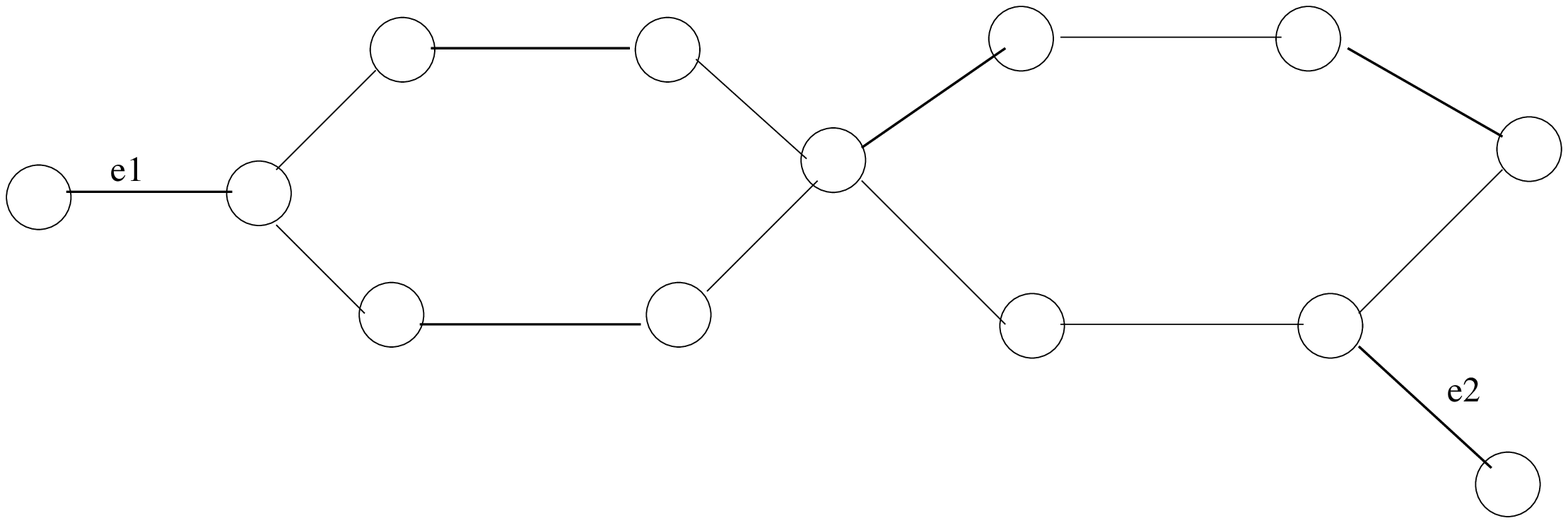}}
\label{fig4}
\caption{Any matching of $G(V,E-\{e_2\}$ is not uniquely restricted }
\end{figure}
It only remove edges $e_2$, then any maximum matching is not uniquely restricted.  
Therefore, greedy algorithms on removing vertex with degree $1$ could not efficiently to determing all 
maximum matching are unqiuely restricted.
 
However, a uniquley restricted maximum matching will include vertex with degree $1$, let us define a matching $M$ is a greedy matching if the free vertex set $V_f$ of $M$  do not include vertex with degree  or  all edges of  $G[M]$  saturate a vertex degree $1$.
Then let us define an extend BD-mapping digraph, which consist of  the free vertex set $V_f$ and a  BD-mapping digraph $D$ , where $M$ is greedy matching.


\begin{defn}
\label{extend_RZ_Mapping} 
An extend BD-mapping digraph of bipartitie graphs $G(V,E)$ with  a greedy matching $M$ is follows.  $V= V_f \cup V_1$  and $A=(A_1 \cup  A_f)$, where  $D(V_1,A_1)$ is a RZ-mapping digraph and $A_f$ is a pair of  arcs between $(v_i, v_j) \in G$($ v_i \in V_f$),  and all of $v_i \in M$  if $d(v_i)=1$
\end{defn}

An example of extend $BD-mapping$ is shown in Fig.$5$. Now let we give a necessary and sufficient condition  of all maximum matching are uniquely restricted.
\begin{figure}[htbp]
\centerline{\includegraphics[scale=1]{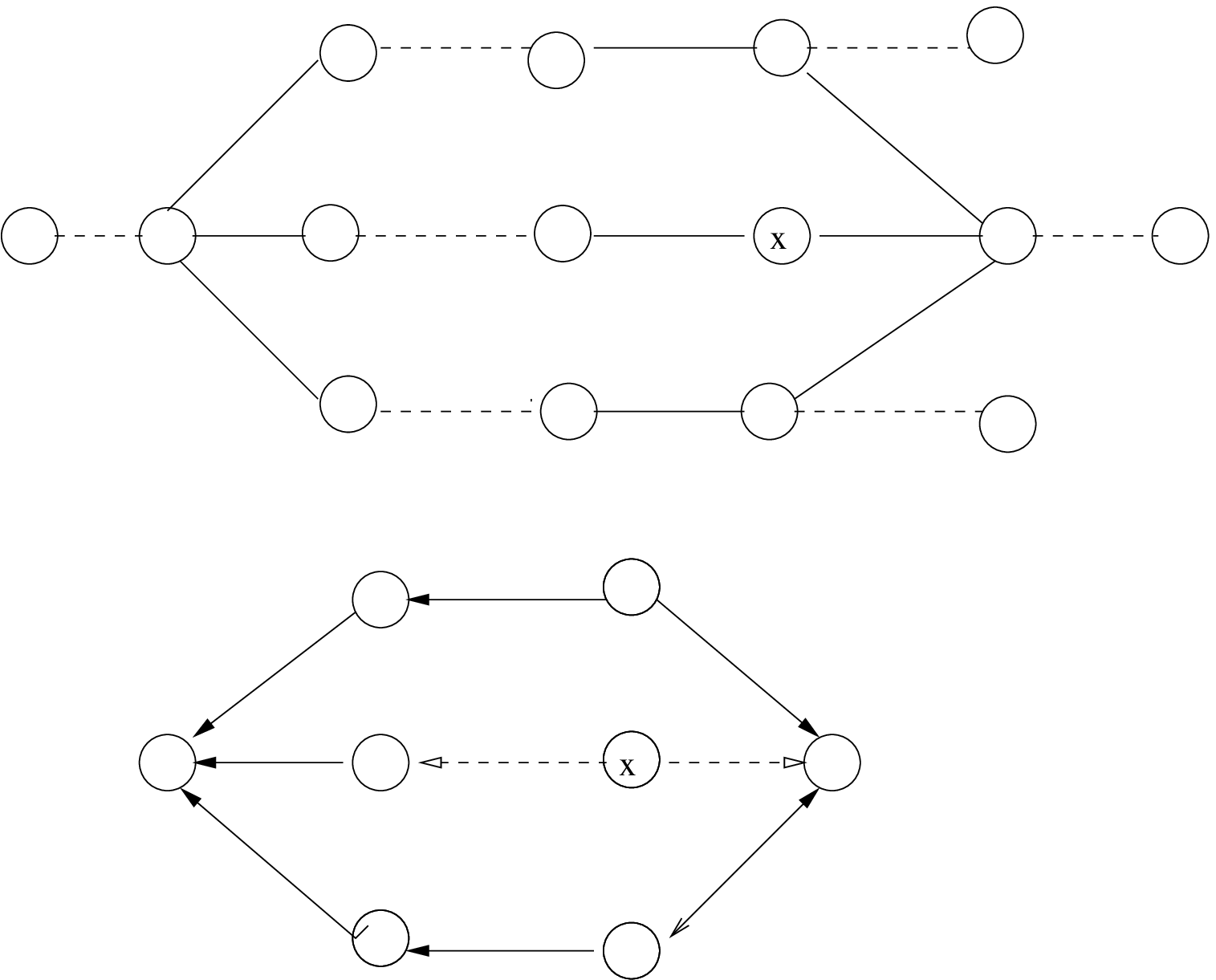}}
\label{fig5}
\caption{Extend BD-mapping from a matching of graph to a digraph }
\end{figure}

\begin{thm}
\label{main_thm2}
 Let $D$ response to the extends BD-mapping digraph of a bipartite graph $G$ with uniquely restricted maximum matching and $V_t$ and $V_s$ are the set of terminal nodes or start nodes in $D(G,M,V_f)$, when satisfies following one
of three conditions
\begin{description}
\item[c1.] all of $v_i \in V_f$, there exists only one  path from $v_i$ to $v_j \in V_t$.
\item[c2.] all of $v_i \in V_f$, there exits only one path  from  $v_j \in V_s$ to $v_i$.
\item[c3.] for any two $v_i,v_j \in V_f$, if there exists at most  one $v_k \in D(G,M)$  have the path from $v_i$ to $v_k$ and  $v_j$ to $v_k$, (or conversly, 
there exists at most one $v_k \in D(G,M)$ have the path from $v_k$ to $v_i$ and $v_k$ to $v_j)$.
\end{description}
 then all maximum matching of $G$ are uniquely restricted, where $V_f$ is set of free nodes of $D(G,M)$.
\end{thm}

\begin{pf}
\begin{description}
\item[c1.]
If a free node $v_i \in V_f $ have more than two  disjoint paths $p_1$, $p_2$ to $v_j$,
then  it implies that two consecustive edges $(v_1, v_i)$,$(v_i,v_2)$ not belong to $M$
and  in $P_1 \cup P_2$.

since terminal node $v_j$ in $D$ is always respect to the vertex with  degree $1$.
There exits two disjoint path to $v_j$, which implies that two consecuistive edges
$(v_3,v_j)$, $(v_j,v_4)$ belongs o $M$ and also in $P_1 \cup P_2$. 
Therefor there exists at least 4 edges in a cycle and not in $M$,if we remove the degree 1 node  $v_j$ from $G$,  $M$ minus $1$, but it can plus $1$ by extends the $(v_1,v_i), (v_i,v_2)$ and $(v_3,v_j),(v_j,v_4)$.

\item[c2.]
The  same principle for the start node $v_j$ if $v_j$ is the source node of $D(G,M)$.

\item[c3.]
Let us prove it by constraction, suppose there exits a node $v_i \in V_f$ have the path 
 $P1$ from $v_i$ to $v_{t1}$ and $P2$ from $v_i$ to $v_{t2}$, also there exits a node 
$v_j \in V_f$ have  the path $P3$ from $v_j$ to $v_{t1}$ and $P4$ from  $v_j$ to $v_{t2}$.
Then  there exits a path from $v_i$ to $v_j$ is length of $|P2|+|P4|$( $|P1|+|P3|$ respectively), 
but the  nodes in the $D(G,M)$ is $|P2|+|P4|-2$ ($|P1|+|P3|-2$ respectively), therefor, 
there exists a cycle  in length of $2*(|P1|+|P2|+P3|+|P4|-2)$, which have  a matching  length 
of  $|P1|+|P2|+|P3|+|P4|-2$. 
\end{description}
\end{pf}

According to the theorem~\ref{main_thm2}, it is easy to design a deterministic all maximum matching restrict or not in polynomial algorithm since  deterministic free node $v_i$ to the terminal node
$v_t$ or start node $v_s$  have more than  two disjoint  path is clearly in polynomial time. The Algorithm $2$ shows the algorithms for determine all maximum matching restricted.

\begin{algorithm}
\caption{Determine all maximum matching uniquely restricted  }
\label{alg1}
\begin{center}
    \fbox{
      \begin{minipage}{\textwidth}
      \begin{tabbing}
{\bf Input} : A bipartite graph $G(X,Y; E)$ with a greedy matching $M \in E$; \\
{\bf Output}: return true if all maximum matching are uniquely restricted, otherwise return false.\\
\noindent{\bf begin}	\\
(1).Generate a BD-mapping graph $D(V; A)=f(G[M])$.	\\
(2)  {\bf if }  $M$ is not perfect  then 			\\
(3)  \ \ \ \ \ \ return non unique perfect matching\\
(4)   {\bf    else}   			\\
(5)   \ \ {\bf if} $D$ is acyclic then return unique perfect matching 	\\
(6)   {\bf    else return non unique perfect matching}   			\\
\ \    {\bf endif}	\\
\noindent{\bf end}	
     \end{tabbing}
     \end{minipage}
    }
  \end{center}
\end{algorithm}

\section{Discussion}
The mapping from a matching of bipartite graph to digraph had been succesful solve forcing matching problem in bipartite graph of  \cite{peter2004}\cite{pachter1998}. This paper extends it and use to solve the uniquley restricted maximum matching problem.
According to the theorem~\ref{main_thm2}, the open question appear in \cite{Levit2004} to recognize the all uniquely restricted maximum matching bipartite graphs is solved in polynomial time.





\bibliographystyle{elsarticle-num}
\bibliography{<your-bib-database>}



\end{document}